\documentclass[final,3p, twocolumn,times,review,preprint]{elsarticle}

\usepackage{amssymb}
\usepackage{xcolor}
\usepackage{lipsum}
\usepackage{subfigure}
\usepackage{subcaption}
\usepackage{amsmath}

\journal{Journal of Nuclear Materials}

\begin{document}

\begin{frontmatter}




\title{Defects and Impurity Properties of VN Precipitates in ARAFM Steels: \\ Modelling using a Universal Machine Learning Potential and Experimental Validation}

\author[ICL]{R. S. Stroud}
\author[ICL]{C. Reynolds}
\author[UKAEA]{T. Melichar}
\author[UKAEA]{J. Haley}
\author[OX]{M. Carter}
\author[OX]{M. Moody}
\author[UKAEA]{C. Hardie}
\author[UKAEA]{D. Bowden}
\author[UKAEA,OX]{D. Nguyen-Manh}
\author[ICL]{M. R. Wenman}

\cortext[cor1]{Corresponding author}

\affiliation[ICL]{organization={Department of Materials and Centre for Nuclear Engineering, Imperial College London},
            addressline={Exhibition Road},
            city={London},
            postcode={SW7 2AZ},
            country={UK}}

\affiliation[UKAEA]{organization={Materials Division, United Kingdom Atomic Energy Authority},
            addressline={Culham Campus},
            city={Abingdon},
            state={Oxfordshire},
            postcode={OX14 3DB},
            country={United Kingdom}}

\affiliation[OX]{organization={Department of Materials, University of Oxford},
            addressline={Parks Road},
            city={Oxford},
            state={Oxfordshire},
            postcode={OX1 3PH},
            country={United Kingdom}}

\begin{abstract}
VN precipitates used to strengthen ARAFM steels for fusion applications dissolve under high Fe ion irradiation (100~dpa at $10^{-3}$~dpa$\cdot$s$^{-1}$, 600$^\circ$C). This study examined point defects and solute substitutions using atom probe tomography, machine learning interatomic potentials, and density functional theory. Combined with transmission electron microscopy, results show N-vacancies and substitutional Cr exist in VN precipitates before irradiation. Monte Carlo simulations and collision cascade simulations confirm ordered vacancies at operating temperatures help mitigate irradiation damage. However, solute additions disrupt vacancy ordering and enhance irradiation-induced damage, potentially accelerating dissolution.
      
\end{abstract}


\begin{highlights}
\item APT results show that Cr, along, with other solutes, are present in these precipitates.
\item The small experimentally measured lattice parameter for VN precipitates can be at least partially explained by a combination of N-vacancies and solutes.
\item NEP and DFT results both predict sub-stoichiometric ground states in the VN system with ordered vacancy layers favoured.
\item Switching from bulk to local reference states changed the apparent stability of vacancies in VN.
\item Monte Carlo and cascades confirm the presence of ordered vacancies at high temperature and their role in mitigating irradiation damage 
\end{highlights}


\end{frontmatter}




\section{Introduction}
One challenge for the commercialisation of fusion energy reactors is the development of reduced activation steels with high resistance to creep and irradiation damage that are easy to manufacture \cite{quadling2024developing}. 
To this end, advanced reduced activation ferritic-martensitic (ARAFM) steel alloys with rocksalt-structured MX phase precipitates, such as vanadium nitride (VN), have been proposed \cite{chen2016effects,liu2012microstructure,huang2013recent, li2024phase}. 
However, Haley et al. recently observed that VN precipitates underwent complete dissolution during a 100 dpa Fe+ ion irradiation \cite{haley2024complete}. 
ARAFM alloys are complex systems with multiple alloying elements in addition to precipitate phases. 
Thus, understanding the thermodynamic factors and role of irradiation damage in precipitate stability is both challenging and necessary.

Although alloy designers have often assumed MX precipitates to have idealized rocksalt stoichiometry and crystal structures, many rocksalt nitrides exhibit a strong preference for vacancy-driven non-stoichiometry at 0 K\cite{ashley2007accommodation,ashley2009mechanisms, lazar2008density, ravi2010cluster, gunda2018first, gunda2018resolving}. 
This preferred vacancy concentration may change if the MX phase is alloyed with other elements, either during precipitation or due to ballistic mixing.
Additionally, during irradiation, the preferred vacancy concentration will be disturbed and other point defects will accumulate, which can further change the stability of the precipitate.
A better understanding of the role of solutes and point defects on the stability and resilience of the VN precipitate under irradiation and different temperatures is needed.

We report a combined computational and experimental approach that sought to shed light on the structure of VN precipitates in ARAFM steel prior to irradiation and the types of irradiation damage that may cause dissolution. 
Atomic scale modelling, combined with transmission electron microscopy (TEM) and APT results, pointed to a significant concentration of vacancies and solutes in the unirradiated precipitates. 
Due to the high temperature operating conditions of ARAFM steels, we performed Monte Carlo and molecular dynamics at temperature using a finetuned the universal neuroevolution potential (NEP) \cite{fan2021neuroevolution, liang2025nep89}.
The finetuned NEP potential was then used to study the effects of irradiation damage and temperature on stability of the rocksalt V-N-X (X=Cr, Fe, C, Si, Mn, W, Ta, B, S, P) system.

\section{Methods}
Samples of unirradiated ARAFM steel of the same composition and microstructure, as studied under irradiation by Haley et al, were studied with TEM and APT.
The lattice parameter of VN precipitates could be measured from the series of dark-field images collected by Haley et al. For each dark-field image, a corresponding diffraction pattern was captured, and the position of the objective aperture was recorded. These were compared against a simulated pattern of VN, from which the lattice parameter was deduced.
APT on the unirradiated ARAFM steel was collected via both voltage (20\% pulse fraction, 200 kHz pulse frequency) and laser pulsing methods (40 pJ pulse energy, 200 kHz pulse frequency) at a temperature of 50 K. This data was analysed with APSuite 6.1. The precipitate compositions were measured using proxigrams created from 2.2\% concentration isoconcentration surfaces with a bin size of 0.1 nm over three precipitates.

Density functional theory (DFT) calculations of point defects in VN were carried out using VASP 6 \cite{hafner2008ab,kresse1993ab, kresse1994ab} with the Perdew, Burke, Ernzehof (PBE) exchange correlation functional \cite{perdew1996generalized} with k-point spacing and a planewave cut-off energy set to 0.2$\AA^{-1}$ space and 520 eV, respectively, based on the DFT convergence testing found in the supplementary Figs. \ref{fig:Kspacing} and \ref{fig:Encut}. The projector augmented wave method (PAW) \cite{blochl1994projector, kresse1999ultrasoft} pseudopotentials with semicore p and s states were used for V, Fe, and Cr with Gaussian smearing of 0.1 eV for relaxation followed with a static run using the tetrahedron method with Bl{\"o}chl corrections \cite{blochl1994improved}.
Atomic positions and unit cell parameters of all input structures were allowed to relax to a force convergence of 0.02 $\mathrm{eV} \cdot \AA^{-1}$ per atom.

The thermodynamic stability of the VN and VNX systems was studied by analysing the formation energy crystal structures with point defects and clusters of point defects. Thirty intrinsic point defects were calculated with DFT:  V(N) vacancies, interstitials, and anti-sites. 
An additional 18 DFT calculations were performed for Cr and Fe substitutions on the V sublattice and C substitutions on the N sublattice. 
The ideal rocksalt unit cell has a single unique site for each defect type. 
To vary the defect concentration, a single defect was inserted in rocksalt supercells ranging in size from 8 to 256 atoms.  
Configurations with multiple defects of different types were not considered as the primary objective was to isolate the effect of each defect.
This approach necessarily limits the analysis to non-interacting defects. 

To ensure consistency in the training data, the Materials Project trajectory DFT parameters \cite{Jain2013} were used for training data generation, as the importance of maintaining consistency has been demonstrated by Mazitov et al. \cite{mazitov2025pet}.
The training data consisted of 2797 structures, as detailed in Table \ref{tab:structures}, containing VN, VN doped with solutes, body-centered cubic Fe with solutes and ordered structures from the materials project database \cite{Jain2013}. 
The data was split into a training and testing set at a 80:20 ratio.
In total 11 solutes were included; Fe, B, C, Cr, Mn, P, S, Si, Ta, Ti and W. 
The inclusion of 11 solutes was to enable the subsequent modelling of true chemical complexity found in ARAFM steel.
However, such a large element count will necessarily lower benchmark accuracy against DFT data.
These structures were generated using NepTrainKit \cite{CHEN2025109859} and the Integrated cluster expansion toolkit (ICET). 
The non-ICET enumerated structures had an atomic position rattle of 0.05 \AA\ and a random lattice scaling of 1\%.
All non-enumeration structures were subsampled down by a factor of five using farthest point sampling to ensure training efficiency \cite{eldar1997farthest}. 
The enumeration structures were generated using ICET \cite{aangqvist2019icet} to enumerate all possible configurations of defects and substitutions in the V-N-X system up to 14 atoms for VN and 11 for V-N-X with X denoting one of the 11 solute elements. 

\begin{table*}
\centering
\caption{Summary of DFT training structures}
\label{tab:structures}
\begin{tabular}{lr}
\hline
\textbf{System} & \textbf{Number of structures} \\
\hline
\multicolumn{2}{l}{\textit{VN with solutes}} \\
\quad Structure enumeration (up to 12 atoms) & 1569 \\
\quad Randomly doped VN (32 atoms) & 175 \\
\multicolumn{2}{l}{\textit{VN}} \\
\quad Structure enumeration (up to 14 atoms) & 867 \\
\quad Large VN structures (64 to 96 atoms) & 54 \\
\quad Large VN structures with vacancies (31 to 95 atoms) & 22 \\
\multicolumn{2}{l}{\textit{Other}} \\
\quad Solute reference states in Fe (128 to 129 atoms) & 13 \\
\quad Ordered phases (up to 126 atoms) & 97 \\
\hline
\textbf{Total} & \textbf{2797} \\
\hline
\end{tabular}
\end{table*}

The NEP potential was fine-tuned from the NEP89 foundational model \cite{liang2025nep89}. The potential was fine-tuned with the following training parameters

\begin{equation}
    \lambda_1 = \lambda_2 = 0.2,
\end{equation}
\begin{equation}
    \lambda_E = 10,
\end{equation}
\begin{equation}
    \lambda_F = \lambda_V = 1,
\end{equation}

where $\lambda_1$ and $\lambda_2$ are the $\mathcal{L}_1$ and $\mathcal{L}_2$ norm regularisations respectively.
$\lambda_E$, $\lambda_F$ and $\lambda_V$ are the energy, forces and virials training weights respectively. 
In addition, the training was run for 5000 generations with a population size of 20 and a batch size of 2244, chosen to match the entire training dataset size.
The model performance is presented in Table \ref{tab:metrics}.
The root mean squared error (RMSE) values improved during finetuning and reached comparable errors to large message passing neural networks, such as orb-v3 \cite{rhodes2025orb}, but with a substantially lower inference computational cost of approximately three orders of magnitude.
The testing dataset parity plots for energy, forces, stresses and virials are included in the supplementary information in fig. \ref{fig:training_plots}.
This reduced computational cost was prioritised over only minimal errors to be capable of performing molecular dynamics of large systems.  
Crucially the NEP potential after finetuning was able capture the same vacancy ordered structures on the convex hull as DFT indicating its suitability.

\begin{table}
\centering
\caption{Model performance metrics}
\label{tab:metrics}
\begin{tabular}{lc}
\hline
\textbf{Metric} & \textbf{Error} \\
\hline
RMSE test energy & 0.068 eV \\
RMSE test forces & 0.577 eV/\AA \\
RMSE test virials & 0.263 eV/atom \\
RMSE VN hull distance & 0.242 eV/atom \\
RMSE VN hull distance & 0.093 eV/atom \\
\quad (within 0.2 eV) & \\
\hline
\end{tabular}
\end{table}

Convex hull calculations were performed using the finetuned NEP potential using the atom simulation environment (ASE)\cite{larsen2017atomic} implemented in python.
For these calculations, each structure was relaxed using the Frechet cell filter, which allowed the box and atoms to relax. The fast inertial relaxation engine (FIRE) \cite{bitzek2006structural} was then applied with the convergence criteria of a maximum force component of 0.05 eV$\AA^{-1}$ on each atom or 500 steps. 

Structures were calculated with the NEP as follows. 
First, an ideal conventional unit cell was initialized with ASE.
Supercells were then enumerated using ICET \cite{aangqvist2019icet}. 
The structure enumeration was broken into different series for each defect and substitution type. 
All possible unique structures containing the defect type were enumerated using ICET to a maximum of 14 atoms for the rocksalt structure. The NEP potential was used to calculate formation enthalpies for defect in VN.
The defect types consisted of both V-N anti-sites, both V and N interstitials and both V-N vacancies. 
While this was not an exhaustive search of all phases that could be included in the system, it does include the unique single type defect in the VN rocksalt system and each element dissolved in BCC Fe, i.e. the two phases observed in experimental analysis.

Equation \ref{Eq:Hf} shows an example calculation of the enthalpy of formation, $\Delta H_f$, for a VN structure with $n$ unit cells and interstitial solute $S$. 
\begin{equation}
        \Delta H_f = H_{(VN)_{n}S}- (H_{(VN)_n} +(H_{Fe_xS}-H_{Fe_X}))
        \label{Eq:Hf}
\end{equation}
While formation energies are typically calculated relative to the elements standard crystal structures, we have chosen to use the energy of elements as solutes in (BCC) iron.
This was chosen because we are most interested in the stability of these compounds when they are embedded in the steel matrix.
Prior APT results had shown no V-containing precipitates other than VN before or after irradiation, which motivates the use of the energy of V and N dissolved in an Fe matrix as the reference energies.
Their formation and dissolution will depend on the surrounding nearby solutes rather than pure elemental materials. 
The reference structures for the interstitial-type elements (C and N) consisted of 128 Fe atoms and a single interstitial atom. Both the tetrahedral and octahedral site calculations were performed, and the lowest-energy sites were selected.

To investigate the ordering of vacancies at the high temperatures targeted by ARAFM steel of around 900 K an on-lattice Monte Carlo approach was taken to address the 0 K limitation of DFT.
A cubic supercells containing 2744 atoms were used with different initial compositions of VN with solutes and vacancies.
The vacancies are represented as a unique species labelled X. 
A Monte Carlo algorithm was then used to find a representative low energy configuration for each of the given conditions.
The simulations were carried out on-lattice to enable the swapping of vacancies, which remains difficult for off-lattice simulations. 

The Monte Carlo algorithm used is as follows. Where $E_i$ is the energy before a swap, $E_f$ is the energy after a swap, $k_B$ is the Boltzmann constant and $T$ is temperature.

\begin{equation}
\Delta E = E_f - E_i
\end{equation}

\begin{equation}
\begin{aligned}
\text{Decision} &= 
\begin{cases}
\text{Accept swap,} & \text{if } \Delta E < 0 \text{ or } \xi < \exp\left(-\frac{\Delta E}{k_B T}\right) \\
\text{Reject swap,} & \text{otherwise}
\end{cases} \\
&\text{where } \xi \text{ is a uniform random variable on } [0,1]
\end{aligned}
\end{equation}

To carry out the Monte Carlo simulations the software package Clease \cite{chang2019clease} was used with the finetuned NEP potential.
The simulations were run for a total of 70,000 attempted swaps for 3 conditions; VN at 933 K with different initial vacancy concentrations, VN with 25\% vacancies on the N sublattice at temperatures between 0 and 2000 K and VN with solutes and vacancies representative of the APT concentration of 5\% vacancy and 5\% C substitutions on the N sublattice and 12\% Cr, 1\% Mn, 1\% W, 2\% Si and 5\% Fe on the V sublattice.   

To quantify the vacancy ordering, the short range order (SRO) parameter was quantified for each neighbour shell, k, from 1 to 10. The SRO was calculated by

\begin{equation}
\alpha_{ij}^{(k)} = 1 - \frac{n_{ij}^{(k)}}{2c_j}
\end{equation}

where $\alpha_{ij}^{(k)}$ is the SRO parameter between elements $i$ and $j$ for neighbour shell $k$, $n_{ij}^{(k)}$ is the number of atoms of type $j$ neighbouring atoms of type $i$ in shell $k$, $c_j$ is the concentration of atoms of type $j$ in the bulk, and the factor of 2 scales the SRO to one sublattice.

The final structures from the following Monte Carlo runs at 933 K; (5\%, 14\%, 26\%, 50\%) vacancy concentration and the APT representative concentration were used as the starting configurations for primary knock-on atom (PKA) cascades.
The PKA cascades were carried out three times for each of the [100], [110] and [111] directions and for each starting structure.
The PKA energy was set to 1200 eV and the temperature of the simulation was set to 900 K.
The structure were first relaxed using the FIRE algorithm for 2000 steps and then thermalised using the stochastic rescaling NPT ensemble \cite{bernetti2020pressure} for between 5 ps to 14 ps.
The thermalisation time was varied between each run of the same structure to result in a different random seed for the cascade simulation.
After thermalisation, the central N atom was selected and given a kinetic energy of 1200 eV.
The NVE ensemble was used in the central region and a Nose-Hoover chain (NHC) thermostat was set to a 10 \AA{} thick boundary at the outer edge of the supercell.
The NHC thermostat was given a target temperature of 900 K with a temperature coupling constant of 200 K.
The timestep was set at 1 fs with a maximum displacement per step of 0.015 \AA. 

To quantify the damage following the cascades the change in stored energy was calculated. 
This was used instead of the defect count that would be unreliable in this case due to the presence of vacancy defects in the initial structures.
The stored energy, $\Delta E_\mathrm{stored}$, is given by the relaxed energy after the cascade, $E_\mathrm{final}$ minus the relaxed energy before the cascade $E_\mathrm{initial}$, or simply
\begin{equation}
    \Delta E_\mathrm{stored} = E_\mathrm{final} - E_\mathrm{initial}.
\end{equation}

\section{Results}

\begin{figure*}
    \centering
    \includegraphics[width=1\linewidth]{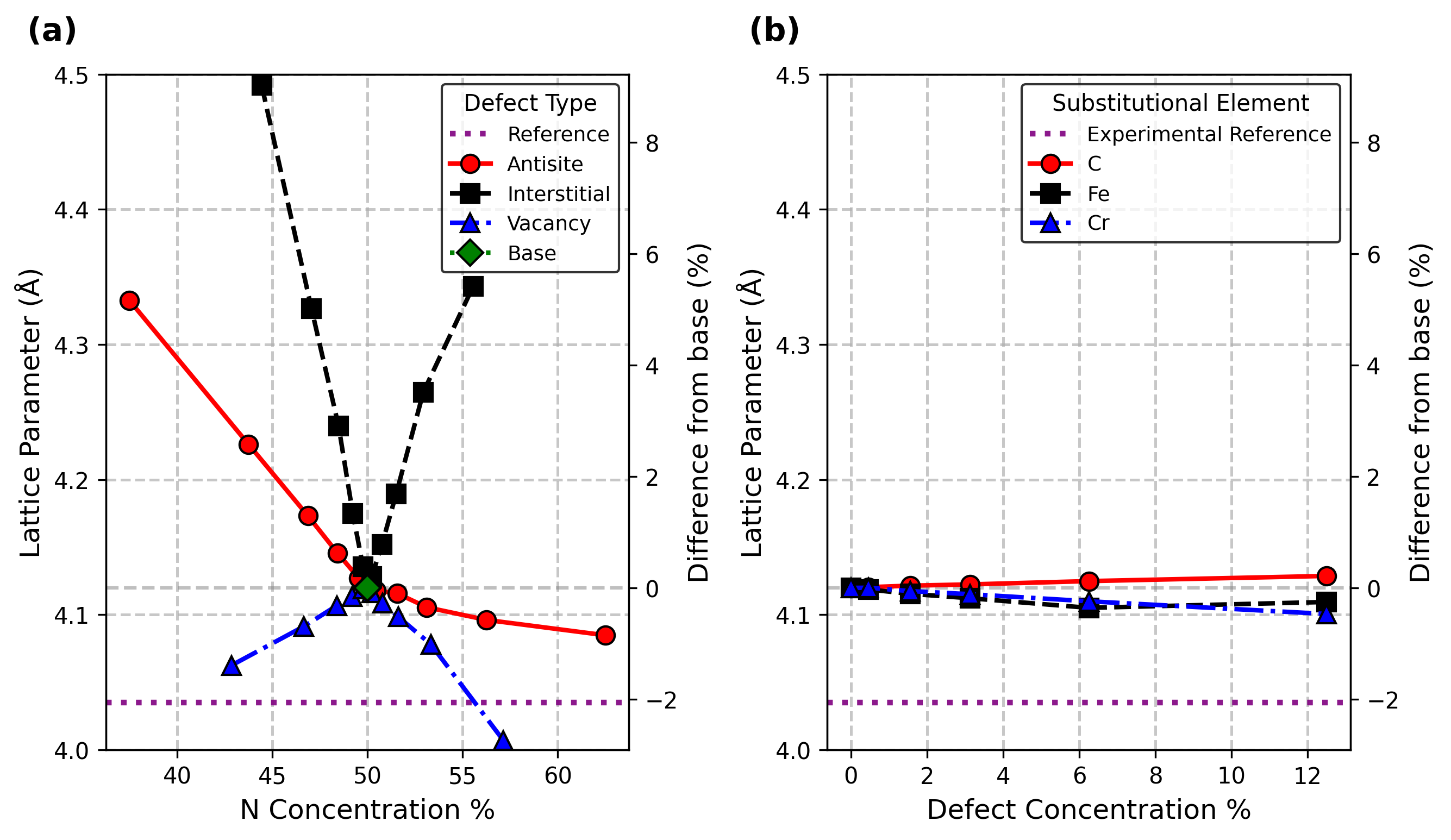}
    \caption{The lattice parameters calculated using VASP for defects in rocksalt VN as a function of concentration are compared to the experimental value (unknown stoichiometry) from TEM highlighted with a horizontal line. The effect of vacancies, interstitial, and antisites are compared in a). The effect of substitutional C, Fe, and Cr on lattice parameter as calculated by DFT are plotted in b) and compared to an experimental value from a single measurement at an unknown stoichiometry (the dashed horizontal line).}
    \label{fig:dft}
\end{figure*}

The experimental lattice parameter for the VN precipitates as measured with TEM was $4.035 \r{A}$, 
a smaller measurement than the previously reported ($4.1287 \r{A}$) \cite{lei2013synthetic} and much smaller than the calculated lattice parameter for the pristine VN rocksalt structure. 
This measurement is compared to the lattice parameter of VN as a function of the concentration of intrinsic point defects (vacancies, antisites, and interstitials) as calculated with DFT in Figure \ref{fig:dft}a). 
Vacancies on both the vanadium or nitrogen sublattices both lead to a reduction in the lattice parameter (Figure \ref{fig:dft}a) of up to 2\%, whereas the anti-sites and interstitials increase the lattice parameter by 4\% and 8\% respectively.

\begin{figure*}
    \centering
    \includegraphics[width=1\linewidth]{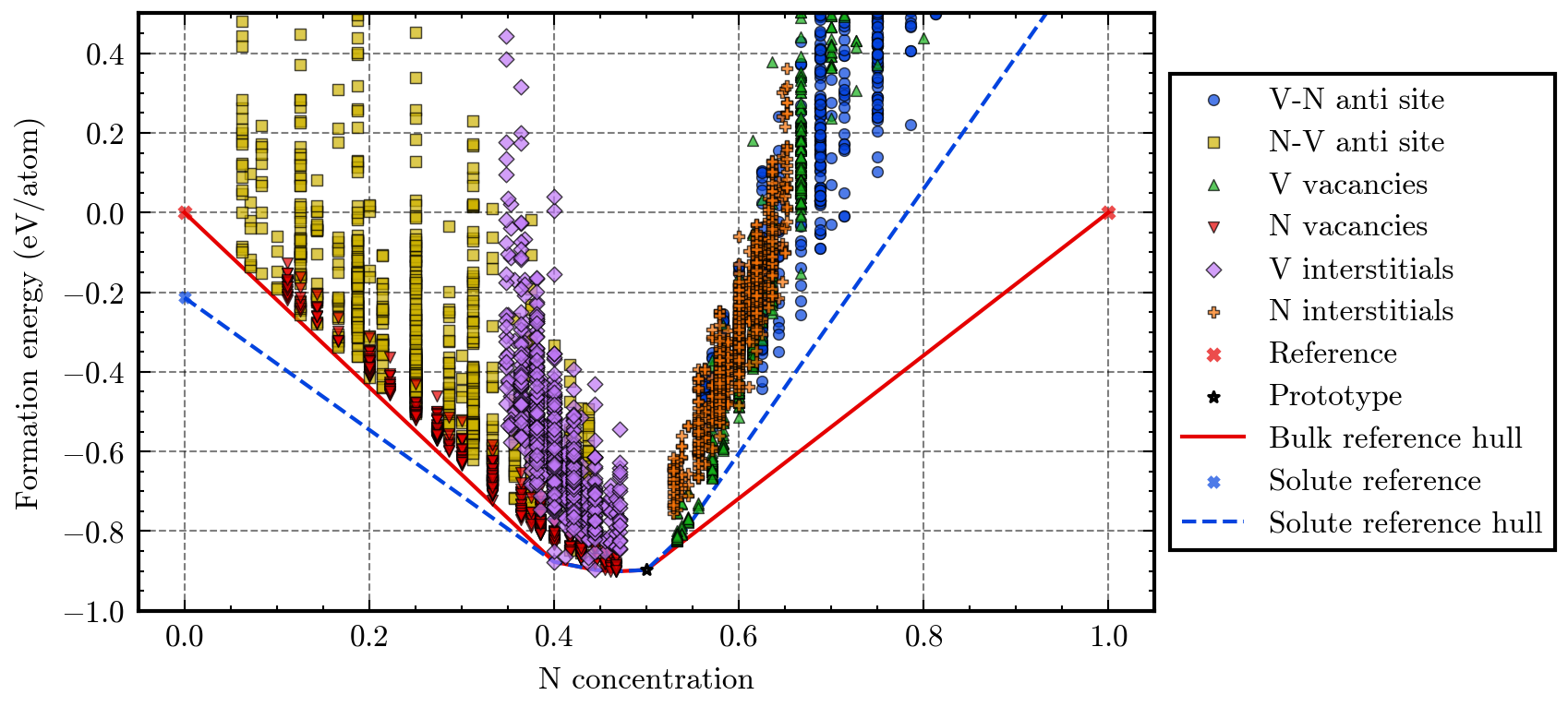}
    \caption{The convex hull calculated with 6519 enumerated structures calculated with the finetuned NEP potential. The red line indicates the convex hull with molecular N and BCC V as reference states, while the blue line is the convex hull where V and N as dilute solutes in Fe are used as reference states.}
    \label{fig:convexhull}
\end{figure*}

The NEP calculated convex hull is shown in Fig \ref{fig:convexhull}. 
This convex hull is in good agreement with a previously reported VN convex hull\cite{ravi2010cluster} in that it reproduces two key features: non-stoichiometric N-vacancy-driven rocksalt structures are predicted as ground states, and these structures have highly ordered layered-vacancy structures.
However, the effect of the chosen reference states plays a role in the relative stability of N vacancies when the VN is considered to be part of the wider system with the solute reference states. 
When the system is referenced to V and N as dilute solutes in Fe, the hull is pulled below the ground states with higher N vacancy concentration. 
Structures with lower N vacancy concentrations remain on the hull. 
These calculations combined with the observed contracted lattice parameter suggest that N vacancies may be present in the VN precipitates in significant concentrations prior to irradiation.

\begin{figure}
    \centering
    \includegraphics[width=1\linewidth]{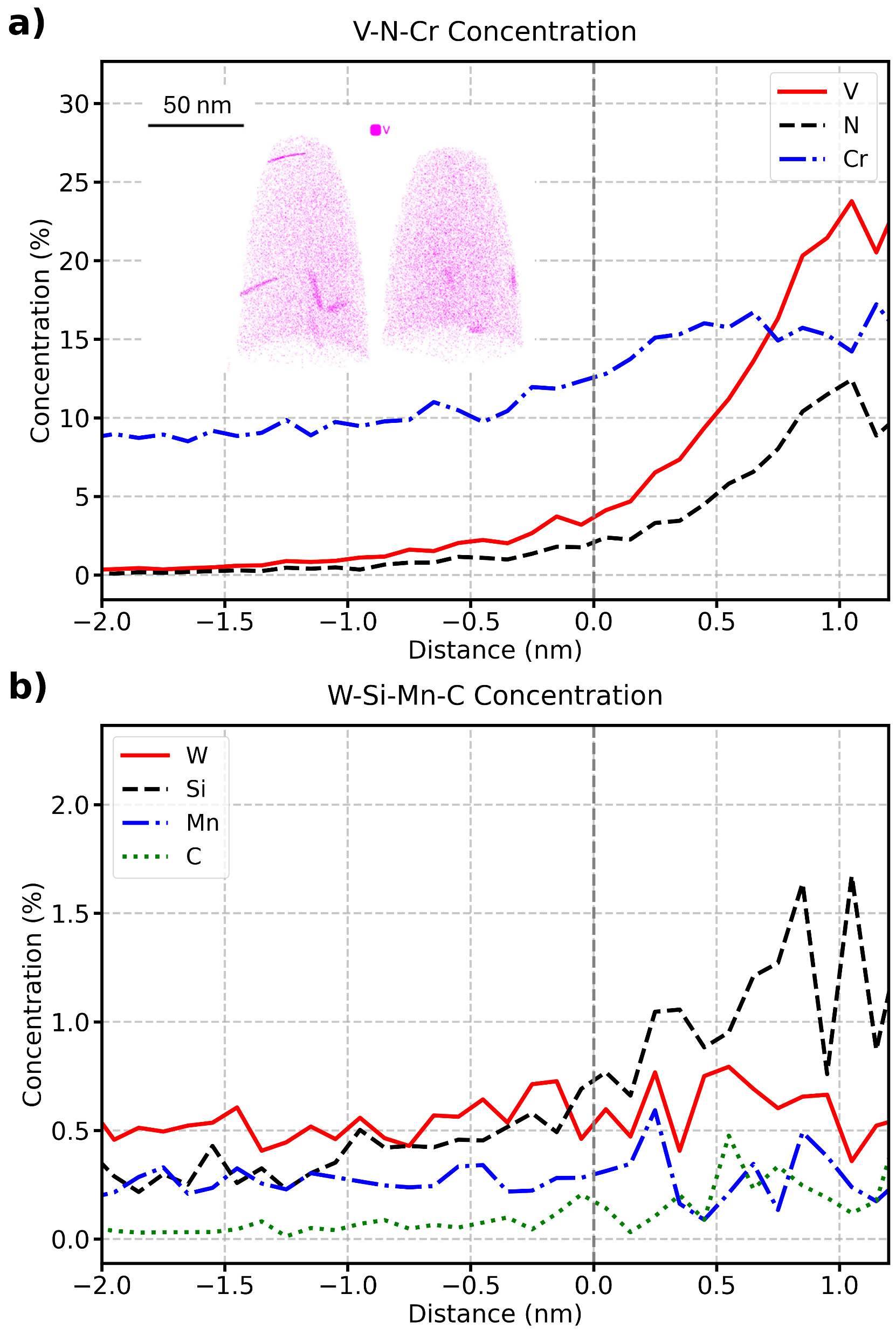}
    \caption{a) The concentration profiles of decomposed V, N, and Cr through the interface of a 2.2\% isoconcentration surface of an APT reconstruction averaged over 3 precipitates. b) The concentration profiles of decomposed C, Si, W and Mn through the interface of a 2.2\% isoconcentration surface of an APT reconstruction. For both plots, the distance is the distance from the interface of the isosurface, with positive values inside the precipitate and 0.0 nm being the interface marked by a vertical dashed line.}
    \label{fig:combined_APT_Conc}
\end{figure}

Figure \ref{fig:combined_APT_Conc} shows the proxigrams of V, N, Cr, Si, W, Mn, and C across all interfaces from RAFM steel matrix and VN precipitate in a unirradiated sample. 
Cr, Si, W, Mn, and C were all present inside the precipitates with Cr and Si concentrations increasing towards the centre of the precipitates.
The vanadium concentration is approximately double that of the measured nitrogen concentration, Figure \ref{fig:combined_APT_Conc} a), though this is likely exaggerated due to an under-measurement of N by APT.
Some N type vacancies may be present as previously discussed, though the concentration is likely less pronounced than this measurement suggests.

DFT calculations of the lattice parameters as a function solute concentration are compared to the observed lattice parameters in Fig. \ref{fig:dft} b). 
Substitutions of carbon on the nitrogen sublattice increase the lattice parameters while both Fe and Cr substitutions on the vanadium sublattice are predicted to decrease the lattice parameter. Although substitutions change the lattice parameter, the effect is minimal with changes being less than 0.2\%.
Thus, Cr and Fe substitutions may slightly alter the lattice parameter but neither are a significant driver of the lattice contraction measured here.

\begin{figure*}
    \centering
    \includegraphics[width=1\linewidth]{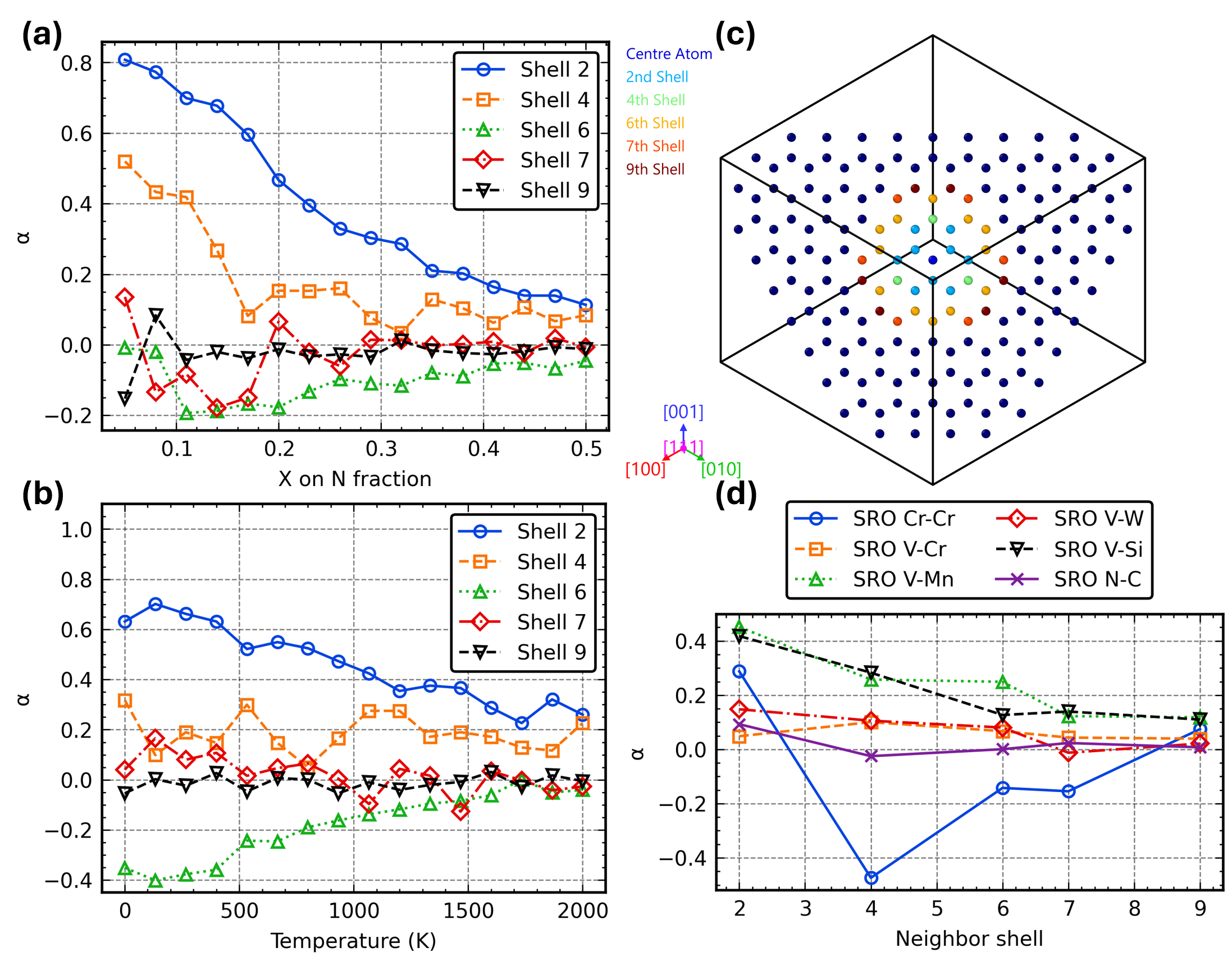}
    \caption{The ordering present taken from the final states from on-lattice Monte Carlo simulations. a) shows the SRO parameters as a function of initial vacancies (X) on the N sublattice at a temperature of 933 K. b) shows the SRO parameters as a function of temperature at a vacancy concentration of 25\%. c) an illustration of nearest neighbour shells of N atoms in a slice in the [111] plane viewed down the [111] direction. d) shows the SRO parameters as a nearest neighbour shells including solutes at a temperature of 933 K.  }
    \label{fig:lattice_mc}
\end{figure*}

Ordered vacancy layers have been confirmed in this work at zero temperature and with respect to solute reference state, as shown in the VN convex hull in Fig. \ref{fig:convexhull}. 
However, to investigate the effect of temperature on-lattice Monte Carlo simulations were conducted.
The SRO between vacancy-vacancy at 933 K in the first to tenth nearest neighbour shells (the sites are illustrated in Fig. \ref{fig:lattice_mc} c)) as a function of vacancy concentration was found to trend to zero and random ordering with increasing vacancy concentrations, as shown in Fig. \ref{fig:lattice_mc} a).
In addition, a high SRO parameter can be seen for the shell numbers 2 and 4 highlighting less than random occupations.
The sixth and seventh neighbour shells are found to have the lowest SRO parameters, although the ordering decreases with increasing vacancy concentration.  
The effect of temperature was subsequently investigated with a fixed vacancy concentration of 25\%.
As with the vacancy concentration, increasing temperature was found to decrease ordering, as shown in Fig. \ref{fig:lattice_mc} b). It is noted that significant ordering still exists at 933 K for shells 2 and 6. Shell 4 exhibits unusual behaviour likely a result of fluctuations between many configurations close in energy. 
The APT representative composition was also investigated at 933 K and the SRO as function of shell number is shown in Fig. \ref{fig:lattice_mc} d).
The Cr, Mn, and Si show the most ordering with all solutes having weaker ordering at larger distances.

\begin{figure*}
    \centering
    \includegraphics[width=1\linewidth]{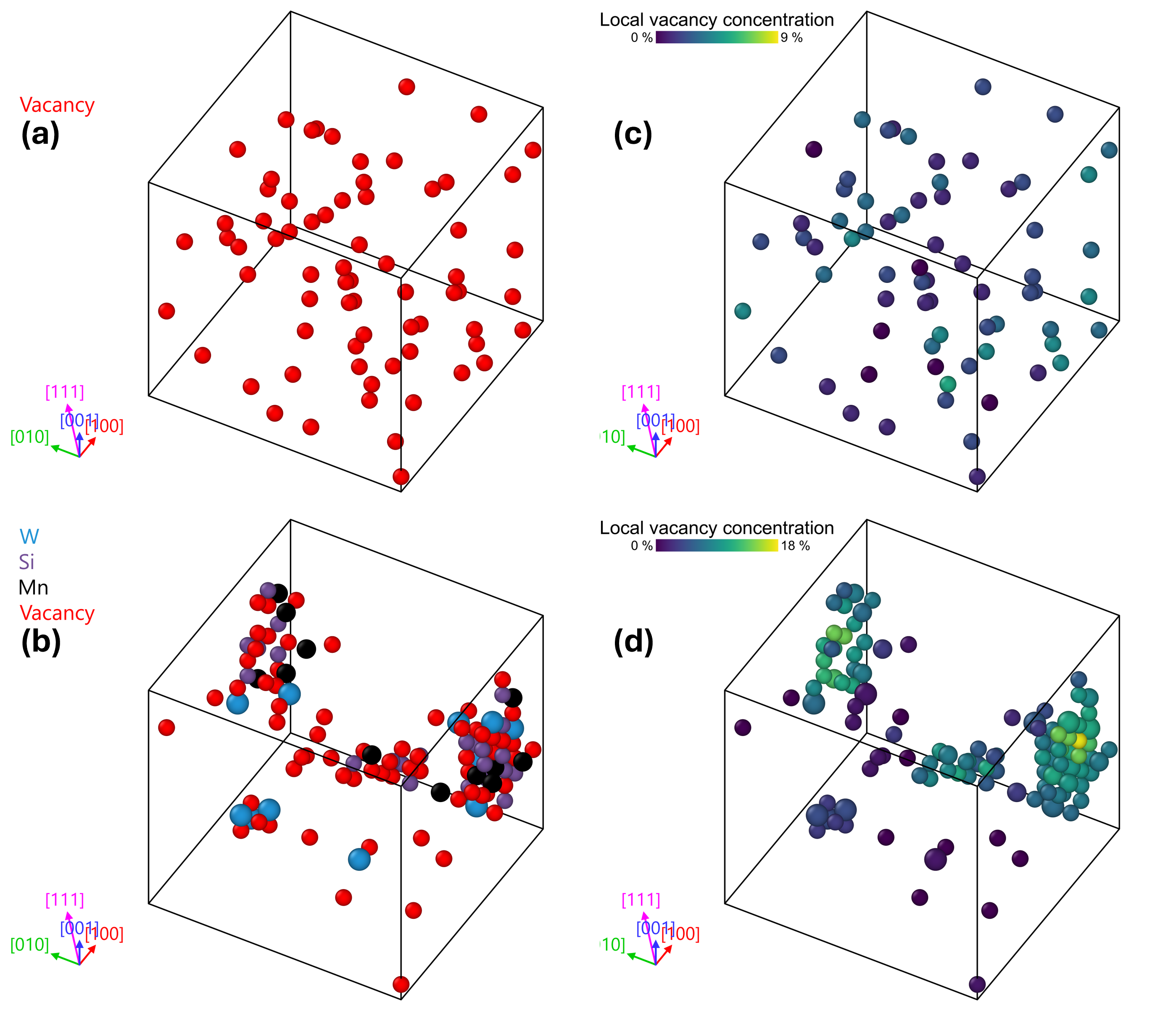}
    \caption{The on-lattice Monte Carlo atomic configurations taken from the final state of a 5\% vacancy concentration at 933 K with and without solutes. a) and c) show the vacancies in $\mathrm{VN}_{0.95}$ condition coloured by species, a), or by local vacancy concentration (within 3 \AA). b) and d) show the vacancies in $\mathrm{VN}_{0.95}$ containing solutes condition coloured by species, a), or by local vacancy concentration (within 3 \AA).}
    \label{fig:ovito_cluster}
\end{figure*}

The vacancy ordering behaviour showed stark differences with VN and VN with the added solutes. 
The vacancies on the N sublattice has been shown to order along the [111] plane. 
However, with the introduction of solute atoms, most notably large atoms such as W, this ordering is broken down and the formation of clustering like behaviour was observed, as shown in Fig. \ref{fig:ovito_cluster}. 
In particular, in Fig. \ref{fig:ovito_cluster} a) and b) the vacancies form into vacancy-W-Si-Mn clusters. 
This is likely partially because the presence of vacancies mitigating the strain induced from the large W atoms.

\begin{figure}
    \centering
    \includegraphics[width=1\linewidth]{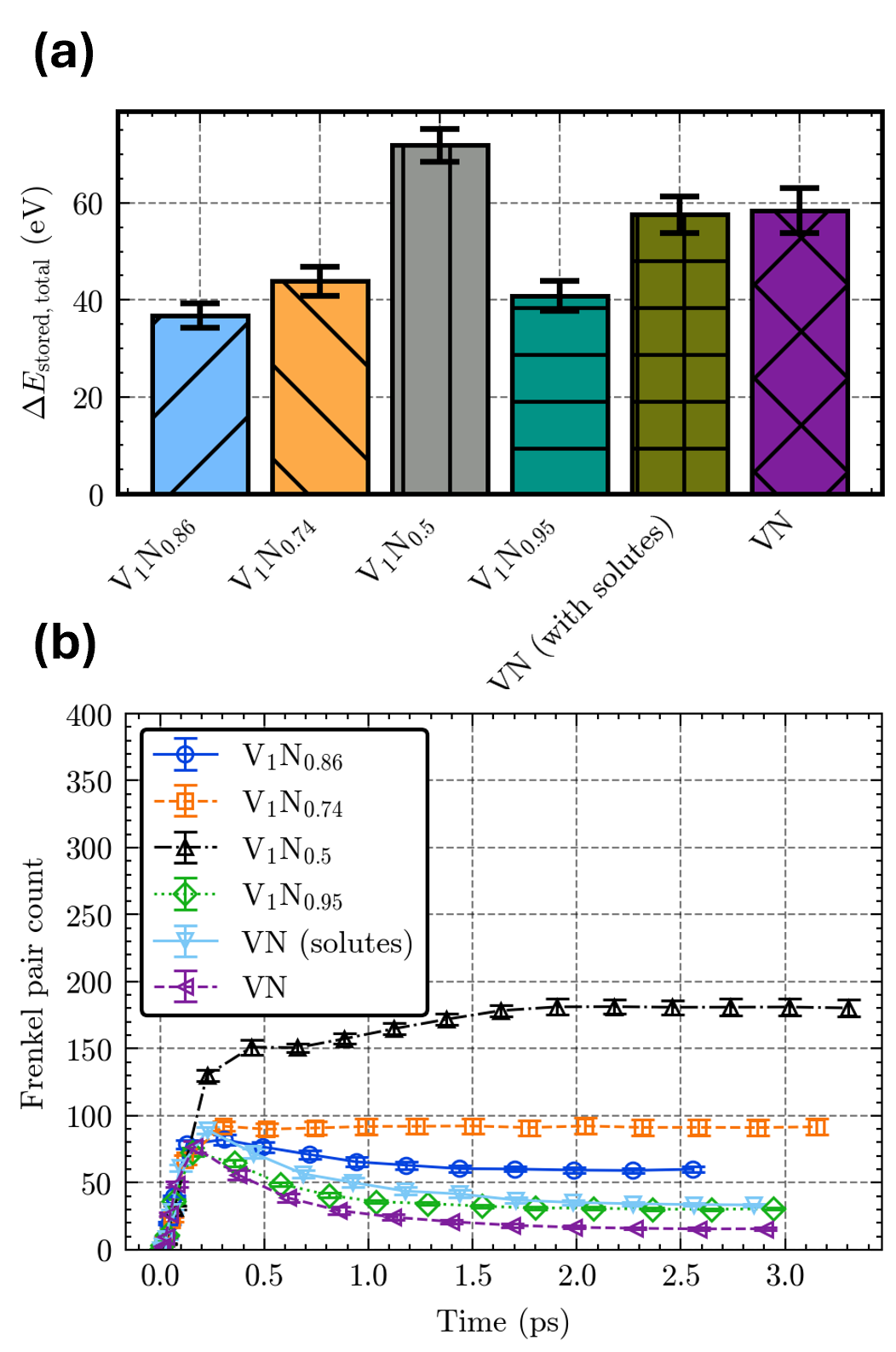}
    \caption{The damage introduced from collision cascades. a) showing the change in stored energy for each condition. b) The Frenkel pair count over time for each condition calculated from the initial frame starting condition.}
    \label{fig:cascade_data}
\end{figure}

To understand the dynamics of irradiation damage, PKA collision cascades were performed on five conditions; VN, VN with a (5\%, 14\%, 26\%, 50\%) pre-existing vacancy concentration and the APT representative concentration. 
All starting structures were taken from the final state of the Monte Carlo simulations. As a result structures containing vacancies started with ordered vacancies, and the APT representative structure contained vacancy-solute clusters.   
The lower initial vacancy concentration conditions of VN$_\mathrm{0.86}$, VN$_\mathrm{0.95}$ and VN$_\mathrm{0.74}$ had a lower increase in stored energy when compared with the pristine VN, as shown in Fig. \ref{fig:cascade_data} a). 
The higher vacancy concentration VN$_\mathrm{0.5}$ was found to have a larger increase in stored energy over the pristine case.
Finally, the APT composition had an apparent protective effect due to the presence of 5\% vacancies and an apparent detrimental effect due to the addition of solutes.
The two competing effects result in a comparable stored energy to the pristine condition. The Frenkel pair count over time, given by Fig. \ref{fig:cascade_data} b), shows a clear relationship between vacancy number and final number of Frenkel pairs. However, this is caused by the measurement including the movement of defect as opposed the their creation. Therefore, the stored energy is a more applicable metric that include the increase in energy from both the disordering of defects as well as the creation of new defects.  

\section{Discussion}
APT results show that VN precipitates in these ARAFM may not be modelled appropriately as only a V-N system and that ternary or more complex systems may be required. 
However, DFT, whilst commonly used for its accuracy, is too computationally expensive to study such systems. 
Here, the NEP89 potential was finetuned and employed to explore a larger systems that cannot easily be explored with DFT alone, and it can do so capably.
The potential was able to replicate DFT-level results for stable VN structures \cite{ravi2010cluster}, including finding previously reported stable vacancy layers.  

Experimentally, the pre-irradiation vacancy and solute concentration for the VN precipitates is difficult to measure due to the small precipitate size and the limitations of APT. The point defect calculations for VN show a preference for vacancy-driven non-stoichiometry with a thermodynamic preference for layered vacancy structures proposed by \cite{gunda2018first, gunda2018resolving}. However, when accounting for the local environment by adjusting the reference states the preference for N vacancies was weaker than previously reported reducing the range of stable stoichiometries present, highlighting a need to reconsider MX-type precipitates in the context of the surrounding steel matrix. The low vacancy concentration structures (5\% to 50\%) were still present on the convex hull. Including N vacancies and Cr-V substitutions may partially explain the small experimental lattice parameter of 4.035 \AA{}, although the Cr had little effect. 

Both this work and prior work in the literature\cite{ravan2019ab} indicate that VN can accommodate N vacancies, with highly ordered layered vacancy structures as ground states.
However, irradiation damage at 10s of dpa is expected to destroy long-range ordering in these precipitates. The particular temperature and fluence experienced by the material could result in a dynamic equilibrium by which irradiation destroys ordered structures generating randomly ordered vacancies, and a thermodynamic driving force drives diffusion to reorder the structure into vacancy layers.
It is shown here that there remains ordering of vacancies at the target operating temperature of over 800 K, albeit the ordering is reduced.  
In addition, the effect of single cascades suggests that low to moderate vacancy concentrations of (5\% to 50\%) may be more robust to irradiation damage on account of deposited incoming energy leading to the reordering of defects in addition to creating them.
Lastly, the effect of solutes seems to play a major role in irradiation damage of VN, and as such caution should considered when such solutes are excluded.
In particular, solutes act to disrupt the vacancy ordering on account of introducing strain into the lattice.
Such strain leads to the formation of solute plus vacancy clusters and is shown with cascades (Fig. \ref{fig:cascade_data}) to increase the stored energy relative to the 5\% vacancy condition. Whilst the lattice Monte Carlo model showed vacancy-solute cluster formation, it is important to recognise the model does not account for kinetics. 
It is known that cluster composition and formation depends partially on the diffusion rates of solutes and defects.
In particular, it is found here that W is a key component of solute clustering in VN, although given its large atomic radii it is expected to diffuse slowly.
A model accounting for both the energetics and kinetics may not find the same vacancy behaviour.

\section{Conclusions}
A combined experimental and computational approach have led to both a better understanding of the VN precipitates in ARAFM steel and the effect of common irradiation defects on precipitate stability. These findings can be summarized as follows.
\begin{itemize}
    \item A contracted lattice parameter measurement and APT results show that Cr, along, with other solutes, are present in these precipitates, suggesting the VN precipitates should not be modelled as an ideal binary rocksalt phase.
    \item The small experimentally measured lattice parameter for VN precipitates can be at least partially explained by a combination of N-vacancies and solutes. Stoichiometric vacancies on both sublattices may also contribute to the change in lattice parameter, though this was not explored in this study.
    \item NEP and DFT results both predict sub-stoichiometric ground states in the VN system with ordered vacancy layers favoured over disordered point vacancies.
    \item Switching from bulk to local reference states changed the apparent stability of vacancies in VN. This highlights how a calculated binary phase diagram may lead to inaccurate expectations when predicting phase stability of precipitates in more complex alloy systems, such as MX precipitates in ARAFM steel. Reconsidering the reference states is a simple way to gauge phase stability in these alloys. Further calculations of the phase stability of other ARAFM steel MX phase precipitates with respect to local reference states are needed. 
    \item Lattice based Monte Carlo confirmed the presence of vacancy ordering at operational temperatures of 800-900+ K in ARAFM steel.
    \item PKA collision cascades showed the importance of vacancies and solutes in stored defect energy introduced from cascades. In particular, the 5, 14 and 26\% vacancy concentrations mitigated damage, whilst the addition of solutes reversed this for the 5\% solute case tested. 
\end{itemize}

\section{CRediT authorship contribution statement}
\textbf{R. S. Stroud}: Conceptualization, data curation, investigation, formal analysis, writing—original draft, and visualization. \textbf{C. Reynolds}: Writing—review and editing and supervision. \textbf{T. Melichar}: Data curation, investigation, formal analysis and visualization. \textbf{J. Haley}: Investigation and writing—review and editing. \textbf{M. Carter}: investigation. \textbf{M. Moody}: Funding acquisition. \textbf{C. Hardie}: Supervision, writing—review and editing. \textbf{D. Bowden}: Supervision. \textbf{D. Nguyen-Manh}: Conceptualization, formal analysis, writing—review and editing and supervision. \textbf{M. R. Wenman}: Conceptualization, formal analysis, writing—review and editing, supervision, and funding acquisition
\section*{Acknowledgements}
We would like to acknowledge computational resources and support provided by the Imperial College Research Computing Service (http://doi.org/10.14469/hpc/2232). We are grateful to the UK Materials and Molecular Modelling Hub for computational resources, which is partially funded by EPSRC (EP/T022213/1, EP/W032260/1 and EP/P020194/1). DNM and TM would like to acknowledge the EUROfusion support for providing access to Marconi-Fusion HPC facility in generation of the DFT data used for this paper. This work has been funded by the NEUtron iRradiatiOn of advaNced stEels (NEURONE) programme via Fusion Futures. As announced by the UK Government in October 2023, Fusion Futures aims to provide holistic support for the development of the fusion sector.

\section*{Supplementary Information}
\clearpage
\begin{figure}
    \centering
    \includegraphics[width=1\linewidth]{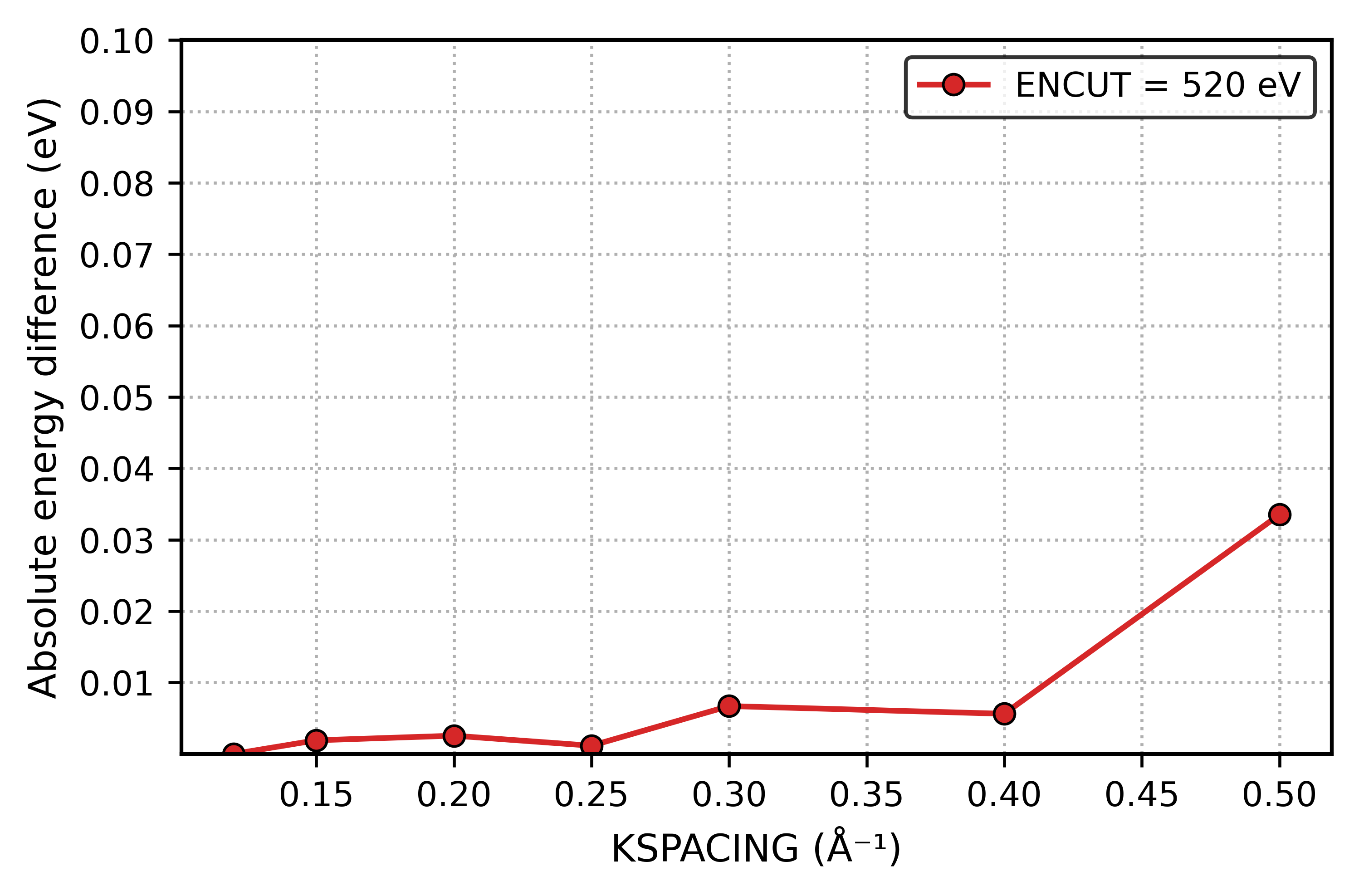}
    \caption{Density functional theory convergence testing for kpoint spacing at 520 eV energy cutoff.}
    \label{fig:Kspacing}
\end{figure}

\begin{figure}
    \centering
    \includegraphics[width=1\linewidth]{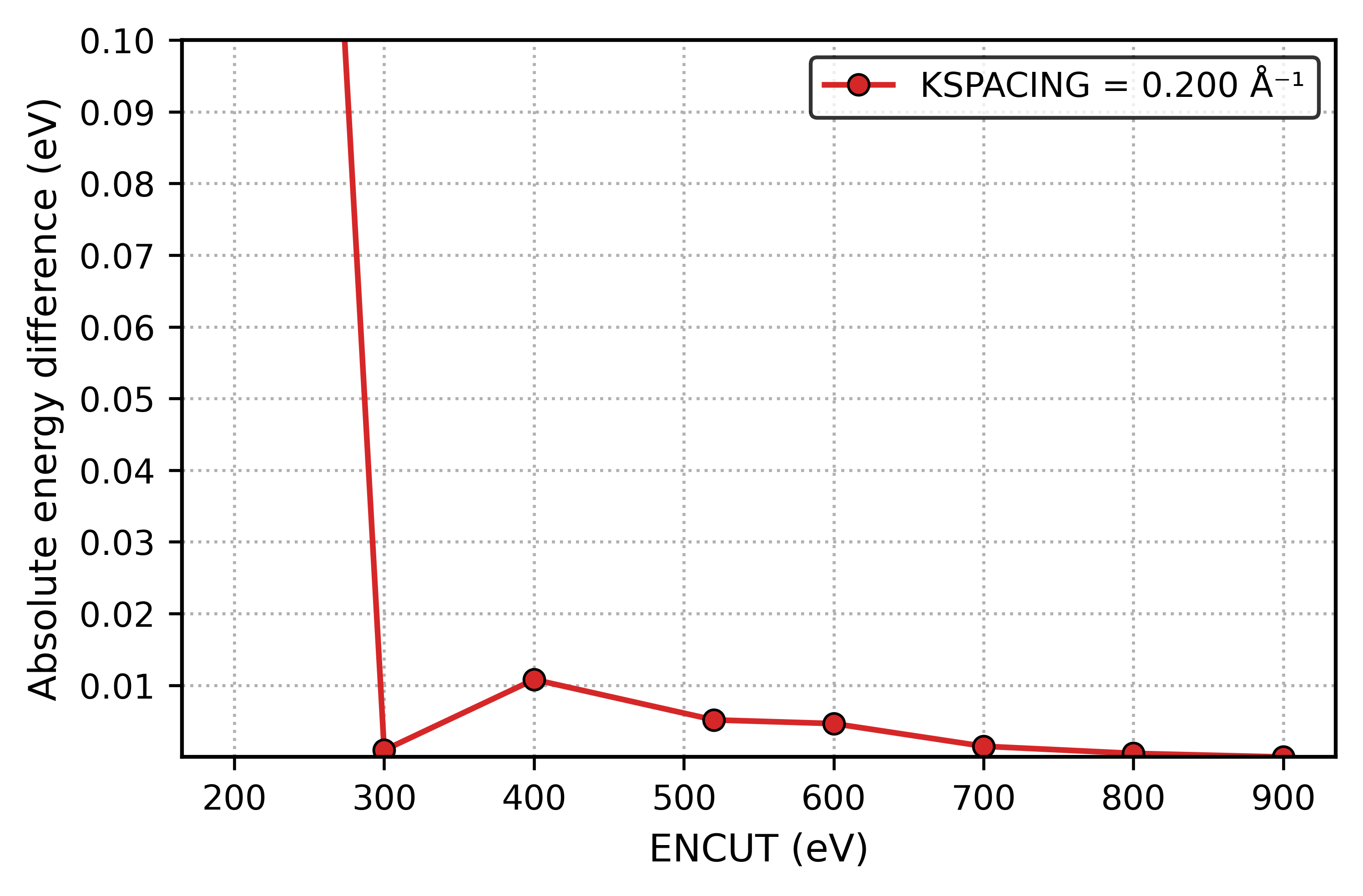}
    \caption{Convergence testing for energy cutoff at a kpoint spacing of 0.2 \AA{}$^{-1}$.}
    \label{fig:Encut}
\end{figure}

\begin{figure*}
    \centering
    \includegraphics[width=1\linewidth]{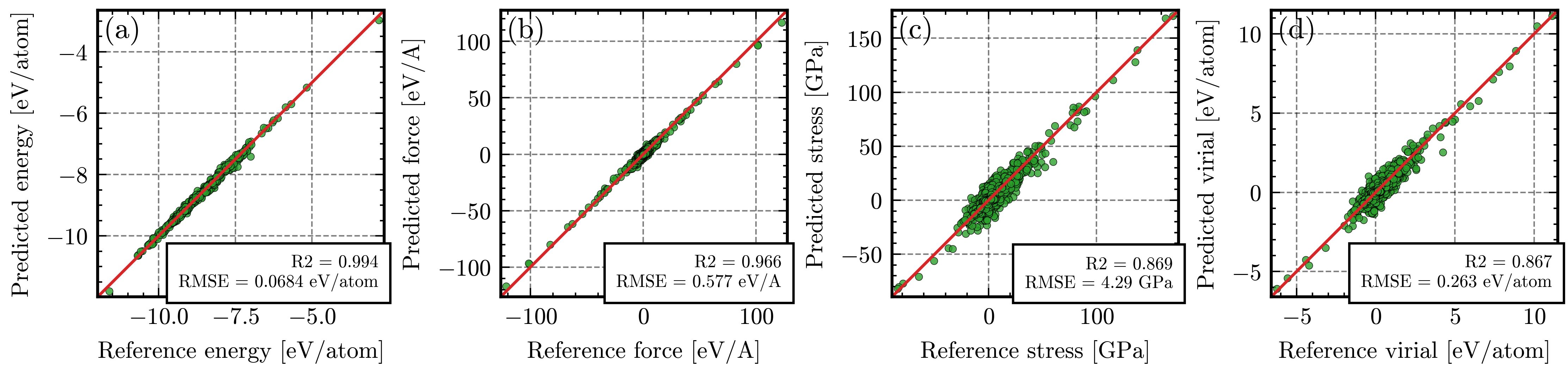}
    \caption{The testing dataset parity plots for the finetuned NEP89 potential against the DFT values for a) energy, b) force, c) stress, and d) virial.}
    \label{fig:training_plots}
\end{figure*}




\clearpage
\bibliographystyle{elsarticle-num}
\bibliography{example}

@article{bernetti2020pressure,
  title={Pressure control using stochastic cell rescaling},
  author={Bernetti, Mattia and Bussi, Giovanni},
  journal={The Journal of Chemical Physics},
  volume={153},
  number={11},
  year={2020},
  publisher={AIP Publishing}
}

@article{chang2019clease,
  title={CLEASE: a versatile and user-friendly implementation of cluster expansion method},
  author={Chang, Jin Hyun and Kleiven, David and Melander, Marko and Akola, Jaakko and Garcia-Lastra, Juan Maria and Vegge, Tejs},
  journal={Journal of Physics: Condensed Matter},
  volume={31},
  number={32},
  pages={325901},
  year={2019},
  publisher={IOP Publishing}
}

@article{rhodes2025orb,
  title={Orb-v3: atomistic simulation at scale},
  author={Rhodes, Benjamin and Vandenhaute, Sander and {\v{S}}imkus, Vaidotas and Gin, James and Godwin, Jonathan and Duignan, Tim and Neumann, Mark},
  journal={arXiv preprint arXiv:2504.06231},
  year={2025}
}

@article{eldar1997farthest,
  title={The farthest point strategy for progressive image sampling},
  author={Eldar, Yuval and Lindenbaum, Michael and Porat, Moshe and Zeevi, Yehoshua Y},
  journal={IEEE transactions on image processing},
  volume={6},
  number={9},
  pages={1305--1315},
  year={1997},
  publisher={IEEE}
}

@article{CHEN2025109859,
title = {NepTrain and NepTrainKit: Automated active learning and visualization toolkit for neuroevolution potentials},
journal = {Computer Physics Communications},
volume = {317},
pages = {109859},
year = {2025},
issn = {0010-4655},
doi = {https://doi.org/10.1016/j.cpc.2025.109859},
url = {https://www.sciencedirect.com/science/article/pii/S0010465525003613},
author = {Chengbing Chen and Yutong Li and Rui Zhao and Zhoulin Liu and Zheyong Fan and Gang Tang and Zhiyong Wang},
}

@article{mazitov2025pet,
  title={PET-MAD, a lightweight universal interatomic potential for advanced materials modeling},
  author={Mazitov, A and Bigi, F and Kellner, M and Pegolo, P and Tisi, D and Fraux, G and Pozdnyakov, SN and Loche, P and Ceriotti, M},
  journal={arXiv preprint arXiv:2503.14118},
  year={2025}
}

@article{Jain2013,
author = {Jain, Anubhav and Ong, Shyue Ping and Hautier, Geoffroy and Chen, Wei and Richards, William Davidson and Dacek, Stephen and Cholia, Shreyas and Gunter, Dan and Skinner, David and Ceder, Gerbrand and Persson, Kristin a.},
doi = {10.1063/1.4812323},
issn = {2166532X},
journal = {APL Materials},
number = {1},
pages = {011002},
title = {{The Materials Project: A materials genome approach to accelerating materials innovation}},
url = {http://link.aip.org/link/AMPADS/v1/i1/p011002/s1\&Agg=doi},
volume = {1},
year = {2013}
}

@article{fan2021neuroevolution,
  title={Neuroevolution machine learning potentials: Combining high accuracy and low cost in atomistic simulations and application to heat transport},
  author={Fan, Zheyong and Zeng, Zezhu and Zhang, Cunzhi and Wang, Yanzhou and Song, Keke and Dong, Haikuan and Chen, Yue and Ala-Nissila, Tapio},
  journal={Physical Review B},
  volume={104},
  number={10},
  pages={104309},
  year={2021},
  publisher={APS}
}

@article{liang2025nep89,
  title={NEP89: Universal neuroevolution potential for inorganic and organic materials across 89 elements},
  author={Liang, Ting and Xu, Ke and Lindgren, Eric and Chen, Zherui and Zhao, Rui and Liu, Jiahui and Berger, Esm{\'e}e and Tang, Benrui and Zhang, Bohan and Wang, Yanzhou and others},
  journal={arXiv preprint arXiv:2504.21286},
  year={2025}
}

@article{lei2013synthetic,
  title={Synthetic route to metal nitrides: high-pressure solid-state metathesis reaction},
  author={Lei, Li and Yin, Wenwen and Jiang, Xiaodong and Lin, Sen and He, Duanwei},
  journal={Inorganic Chemistry},
  volume={52},
  number={23},
  pages={13356--13362},
  year={2013},
  publisher={ACS Publications}
}

@article{hafner2008ab,
  title={Ab-initio simulations of materials using VASP: Density-functional theory and beyond},
  author={Hafner, J{\"u}rgen},
  journal={Journal of computational chemistry},
  volume={29},
  number={13},
  pages={2044--2078},
  year={2008},
  publisher={Wiley Online Library}
}

@article{bitzek2006structural,
  title={Structural relaxation made simple},
  author={Bitzek, Erik and Koskinen, Pekka and G{\"a}hler, Franz and Moseler, Michael and Gumbsch, Peter},
  journal={Physical review letters},
  volume={97},
  number={17},
  pages={170201},
  year={2006},
  publisher={APS}
}

@article{larsen2017atomic,
  title={The atomic simulation environment—a Python library for working with atoms},
  author={Larsen, Ask Hjorth and Mortensen, Jens J{\o}rgen and Blomqvist, Jakob and Castelli, Ivano E and Christensen, Rune and Du{\l}ak, Marcin and Friis, Jesper and Groves, Michael N and Hammer, Bj{\o}rk and Hargus, Cory and others},
  journal={Journal of Physics: Condensed Matter},
  volume={29},
  number={27},
  pages={273002},
  year={2017},
  publisher={IOP Publishing}
}

@article{aangqvist2019icet,
  title={ICET--a Python library for constructing and sampling alloy cluster expansions},
  author={{\AA}ngqvist, Mattias and Mu{\~n}oz, William A and Rahm, J Magnus and Fransson, Erik and Durniak, C{\'e}line and Rozyczko, Piotr and Rod, Thomas H and Erhart, Paul},
  journal={Advanced Theory and Simulations},
  volume={2},
  number={7},
  pages={1900015},
  year={2019},
  publisher={Wiley Online Library}
}

@article{ravi2010cluster,
  title={Cluster expansion Monte Carlo study of phase stability of vanadium nitrides},
  author={Ravi, C and Sahu, HK and Valsakumar, MC and van de Walle, Axel},
  journal={Physical Review B—Condensed Matter and Materials Physics},
  volume={81},
  number={10},
  pages={104111},
  year={2010},
  publisher={APS}
}

@article{lazar2008density,
  title={Density functional theory study of ternary V-Cr-N compounds},
  author={Lazar, Petr and Podloucky, Raimund and Kozeschnik, Ernst and Redinger, Josef},
  journal={Physical Review B—Condensed Matter and Materials Physics},
  volume={78},
  number={13},
  pages={134202},
  year={2008},
  publisher={APS}
}

@article{gunda2018first,
  title={First-principles insights on phase stability of titanium interstitial alloys},
  author={Gunda, NS Harsha and Van der Ven, Anton},
  journal={Physical Review Materials},
  volume={2},
  number={8},
  pages={083602},
  year={2018},
  publisher={APS}
}

@article{gunda2018resolving,
  title={Resolving phase stability in the Ti-O binary with first-principles statistical mechanics methods},
  author={Gunda, NS Harsha and Puchala, Brian and Van der Ven, Anton},
  journal={Physical Review Materials},
  volume={2},
  number={3},
  pages={033604},
  year={2018},
  publisher={APS}
}

@article{haley2024complete,
  title={Complete dissolution of MX-phase nanoprecipitates in fusion steels during irradiation by heavy-ions},
  author={Haley, Jack and Jones, Stephen and Mehraban, Shahin and Lavery, Nicholas and Cullen, Jonathan and Carter, Megan and Moody, Michael and Dawson, Huw and Bowden, David},
  journal={Journal of Nuclear Materials},
  volume={596},
  pages={155115},
  year={2024},
  publisher={Elsevier}
}

@article{ravan2019ab,
  title={Ab initio investigation of mechanical and thermodynamic properties of vanadium-nitride},
  author={Ravan, Bahram Abedi and Faghihnasiri, Mahdi and Jafari, Homayoun},
  journal={Materials Chemistry and Physics},
  volume={228},
  pages={237--243},
  year={2019},
  publisher={Elsevier}
}

@article{ashley2009mechanisms,
  title={Mechanisms of nonstoichiometry in HfN1- x},
  author={Ashley, NJ and Parfitt, D and Chroneos, A and Grimes, RW},
  journal={Journal of Applied Physics},
  volume={106},
  number={8},
  year={2009},
  publisher={AIP Publishing}
}

@article{ashley2007accommodation,
  title={Accommodation of non-stoichiometry in TiN 1-x and ZrN 1-x},
  author={Ashley, Nicholas J and Grimes, Robin W and McClellan, Ken J},
  journal={Journal of materials science},
  volume={42},
  pages={1884--1889},
  year={2007},
  publisher={Springer}
}

@article{chen2016effects,
  title={Effects of tantalum content on the microstructure and mechanical properties of low-carbon RAFM steel},
  author={Chen, Jianguo and Liu, Chenxi and Liu, Yongchang and Yan, Biyu and Li, Huijun},
  journal={Journal of Nuclear Materials},
  volume={479},
  pages={295--301},
  year={2016},
  publisher={Elsevier}
}

@article{huang2013recent,
  title={Recent progress of R\&D activities on reduced activation ferritic/martensitic steels},
  author={Huang, Qunying and Baluc, N and Dai, Y and Jitsukawa, S and Kimura, A and Konys, J and Kurtz, Richard J and Lindau, R and Muroga, T and Odette, George R and others},
  journal={Journal of Nuclear Materials},
  volume={442},
  number={1-3},
  pages={S2--S8},
  year={2013},
  publisher={Elsevier}
}

@article{liu2012microstructure,
  title={Microstructure and its influence on mechanical properties of CLAM steel},
  author={Liu, Shaojun and Huang, Qunying and Peng, Lei and Li, Yanfen and Li, Chunjing},
  journal={Fusion Engineering and Design},
  volume={87},
  number={9},
  pages={1628--1632},
  year={2012},
  publisher={Elsevier}
}

@article{kresse1993ab,
  title={Ab initio molecular dynamics for liquid metals},
  author={Kresse, Georg and Hafner, J{\"u}rgen},
  journal={Physical Review B},
  volume={47},
  number={1},
  pages={558},
  year={1993},
  publisher={APS}
}

@article{kresse1994ab,
  title={Ab initio molecular-dynamics simulation of the liquid-metal--amorphous-semiconductor transition in germanium},
  author={Kresse, Georg and Hafner, J{\"u}rgen},
  journal={Physical Review B},
  volume={49},
  number={20},
  pages={14251},
  year={1994},
  publisher={APS}
}

@article{perdew1996generalized,
  title={Generalized gradient approximation made simple},
  author={Perdew, John P and Burke, Kieron and Ernzerhof, Matthias},
  journal={Physical review letters},
  volume={77},
  number={18},
  pages={3865},
  year={1996},
  publisher={APS}
}

@article{blochl1994improved,
  title={Improved tetrahedron method for Brillouin-zone integrations},
  author={Bl{\"o}chl, Peter E and Jepsen, Ove and Andersen, Ole Krogh},
  journal={Physical Review B},
  volume={49},
  number={23},
  pages={16223},
  year={1994},
  publisher={APS}
}

@article{quadling2024developing,
  title={Developing power plant materials using the life cycle lens},
  author={Quadling, Amanda and Bowden, David and Hardie, Chris and Vasanthakumaran, Arti},
  journal={Philosophical Transactions A},
  volume={382},
  number={2280},
  pages={20230409},
  year={2024},
  publisher={The Royal Society}
}

@article{li2024phase,
  title={Phase stability in the Hf-N and Zr-N systems},
  author={Li, Jonathan and Ober, Derick and Van der Ven, Anton},
  journal={Physical Review Materials},
  volume={8},
  number={12},
  pages={123603},
  year={2024},
  publisher={APS}
}

@article{kresse1999ultrasoft,
  title={From ultrasoft pseudopotentials to the projector augmented-wave method},
  author={Kresse, Georg and Joubert, Daniel},
  journal={Physical review b},
  volume={59},
  number={3},
  pages={1758},
  year={1999},
  publisher={APS}
}

@article{blochl1994projector,
  title={Projector augmented-wave method},
  author={Bl{\"o}chl, Peter E},
  journal={Physical review B},
  volume={50},
  number={24},
  pages={17953},
  year={1994},
  publisher={APS}
}







\end{document}